\documentclass[11pt]{article}

\usepackage{amsmath,amsthm,amssymb,ascmac}
\usepackage{bm}
\usepackage[dvipdfmx]{graphicx}
\usepackage[english]{babel}

\newcommand{\be}{\beta}
\newcommand{\ka}{\kappa}
\newcommand{\bea}{\begin{eqnarray}}
\newcommand{\eea}{\end{eqnarray}}
\newcommand{\aeq}{ &=& }
\newcommand{\aeqe}{ & \equiv & }
\newcommand{\aeqd}{ & \stackrel{\mathrm{def}}{=} & }
\newcommand{\aeqap}{ &\approx & }

\newcommand{\bra}{\langle}
\newcommand{\dbra}{\langle  \hspace{-0.5 mm} \langle}
\newcommand{\ket}{\rangle}
\newcommand{\dket}{\rangle \hspace{-0.5 mm} \rangle }

\newcommand{\mP}{\mathcal{P}}
\newcommand{\mQ}{\mathcal{Q}}
\newcommand{\mO}{\mathcal{O}}
\newcommand{\mL}{\mathcal{L}}

\newcommand{\mN}{\mathcal{N}}

\newcommand{\mD}{\mathcal{D}}

\newcommand{\mJ}{\mathcal{J}}

\newcommand{\mR}{\mathcal{R}}
\newcommand{\mT}{\mathcal{T}}
\newcommand{\mS}{\mathcal{S}}
\newcommand{\mK}{\mathcal{K}}
\newcommand{\tr}{\mbox{Tr}}
\newcommand{\dg}{^\dagger}
\newcommand{\tl}{\tilde}

\newcommand{\defe}{\stackrel{\mathrm{def}}{=}}
\newcommand{\e}{\equiv}
\newcommand{\ep}{\varepsilon}
\newcommand{\ga}{\gamma}
\newcommand{\Ga}{\Gamma}

\newcommand{\al}{\alpha}
\newcommand{\bu}{\bullet}
\newcommand{\sig}{\sigma}

\newcommand{\f}{\frac}
\newcommand{\half}{\frac{1}{2}}
\newcommand{\pr}{\prime}
\newcommand{\dl}{\delta}
\newcommand{\Dl}{\Delta}
\newcommand{\lm}{\lambda}

\newcommand{\om}{\omega}
\newcommand{\Om}{\Omega}

\newcommand{\tot}{{\rm{tot}}}

\newcommand{\st}{{\rm{ss}}}

\newcommand{\la}{\label}
\newcommand{\p}{\partial}
\newcommand{\ke}[1]{ \vert #1 \rangle }
\newcommand{\br}[1]{ \langle #1 \vert }
\newcommand{\abs}[1]{\vert #1 \vert }
\newcommand{\no}{\nonumber}
\newcommand{\hn}{{\hat{n}}}
\newcommand{\ha}{{\hat{\al}}}

\newcommand{\hc}{\mbox{h.c.}}

\newcommand{\re}[1]{(\ref{#1})}
\newcommand{\hs}{\hspace}

\newcommand{\dbr}[1]{ \langle  \hspace{-0.5 mm} \langle #1 \vert }
\newcommand{\dke}[1]{\vert #1 \rangle \hspace{-0.5 mm} \rangle }
\newcommand{\res}[1]{˜\ref{#1}}
\newcommand{\bv}[1]{\big \vert_{#1}}
\newcommand{\Bv}[1]{\Big \vert_{#1}}

\newcommand{\RM}[1]{{\rm{#1}}}

\makeatletter
\@addtoreset{equation}{section}
\makeatother

\def\department#1{\def\@department{#1}}

\title{Excess entropy production in quantum system: Quantum master equation approach}

\author{Satoshi Nakajima\footnote{Email: subarusatosi@gmail.com} \  and Yasuhiro Tokura\footnote{Email: tokura.yasuhiro.ft@u.tsukuba.ac.jp} \\
$Graduate$ $School$ $of$ $Pure$ $and$ $Applied$ $Sciences$, $University$ $of$ $Tsukuba$, \\
${\it 1-1-1}$ , $Tennodai$, $Tsukuba$, ${\it 305-8571}$, $Japan$}

\date{\today}

\begin{document}

\maketitle

\begin{abstract}
For open systems described by the quantum master equation (QME), we investigate the excess entropy production 
under quasistatic operations between nonequilibrium steady states. 
The average entropy production is composed of the time integral of the instantaneous steady entropy production rate and 
the excess entropy production.
We 
propose to 
define average entropy production rate using the average energy and particle currents, which are calculated by using the full counting statistics with QME.
The excess entropy production is given by a line integral in the control parameter space and its integrand is called the Berry-Sinitsyn-Nemenman (BSN) vector. 
In the weakly nonequilibrium regime, we show that BSN vector is described by 
$\ln \breve{\rho}_0$ and $\rho_0$ where $\rho_0$ is the instantaneous steady state of 
the QME and $\breve{\rho}_0$ is that of the QME which is given by reversing the sign of the Lamb shift term. 
If the system Hamiltonian is non-degenerate or the Lamb shift term is negligible, the excess entropy production approximately reduces to the 
difference between the von Neumann entropies of the system. 
Additionally, we point out that the expression of the entropy production obtained in the classical Markov jump process is different from our result 
and show that these are approximately equivalent only in the weakly nonequilibrium regime.

\end{abstract}

\section{Introduction} \la{Introduction}

In equilibrium thermodynamics, the central quantity is the entropy $S$, which describes both the macroscopic properties of equilibrium systems and
the fundamental limits on the possible transitions among equilibrium states. 
In equilibrium thermodynamics, the Clausius equality
\bea
\Dl S =\be Q,
\eea
tells us how one can determine the entropy by measuring the heat. 
Here, $\Dl S$ is the change in the entropy of the system during the operation, $\be$ is the inverse
temperature of the bath contacting with the system, and $Q$ is the heat transferred from the bath to the system during the operation. 
This equality is universally valid for quasistatic transitions between equilibrium states. 
In the equilibrium classical (quantum) system, the entropy is given by the Shannon entropy of the probability distribution 
(von Neumann entropy of the density matrix) of states. 

The investigation of thermodynamic structures of nonequilibrium steady states (NESSs) has been a topic of active research in nonequilibrium statistical mechanics 
\cite{18,Oono,SasaTasaki,S10,S11,S12,K17,K33,K35}. 
For instance, the extension of the relations in equilibrium thermodynamics, such as the Clausius equality, to NESSs has been one of the central subjects. 
Recently there has been progress in the extension of the Clausius equality to NESSs \cite{Komatsu08,Komatsu11,Saitou} (see also Refs.\cite{22,5,6,Hatano,Sasa14,K13}).
In these studies, the excess heat $Q_{b,\rm{ex}}$ (of the bath $b$) \cite{Oono}
has been introduced instead of the total heat $Q_b$ from the bath $b$. 
The excess heat $Q_{b,\rm{ex}}$, which describes an additional heat induced by a transition between NESSs with time-dependent external
control parameters, 
is defined by subtracting from $Q_b$ the time integral of the instantaneous steady heat current from the bath $b$. 
In the weakly nonequilibrium regime, there exists a scalar potential $\mS$ of the control parameter space which approximately satisfies the {\it extended Clausius equality}
\bea
\sum_b \be_b Q_{b,\rm{ex}} \approx \Dl \mS. \la{2}
\eea
Here, $\be_b$ is the inverse temperature of the bath $b$, $\Dl \mS=\mS(\al_{\tau})-\mS(\al_{0})$, 
$\al_t=(\al_t^1,\al_t^2,\ldots)$ is the value of the set of the control parameters at time $t$, 
and $t=0$ and $t=\tau$ are the initial and final times of the operation. 
We assume multiple baths to maintain the system out of the
equilibrium and the symbol $\sum_b$ means the summation over the baths.
In classical systems, $\mS$ is the symmetrized Shannon entropy \cite{Komatsu11}. 
In quantum systems with the time-reversal symmetry, $\mS$ is the symmetrized von Neumann entropy \cite{Saitou}.
However, the study of the excess entropy in the quantum system without the time-reversal symmetry seems still lacking.
This is the main objective of this paper.

In general, the left-hand side (LHS) of \re{2} is replaced by the excess entropy 
\bea
\sig_\RM{ex} \defe \sig-\int_{0}^{\tau} dt\ J^\sig_\st(\al_t),
\eea
where $\sig$ is the average entropy production and $J^\sig_\st(\al_t)$ is the instantaneous steady entropy production rate 
at time $t$ \cite{Sagawa,Yuge13,Komatsu15}. 
In the quasistatic operation, the excess entropy is given by  
\bea
\sig_\RM{ex} =\Dl \mS+\mO(\ep^2 \dl),
\eea
where $\ep$ is a measure of degree of nonequilibrium and $\dl$ describes the amplitude of the change of the control parameters.
Sagawa and Hayakawa \cite{Sagawa} studied the full counting statistics (FCS) of the entropy production for classical systems described by the Markov jump process and showed that
the excess entropy is characterized by the Berry-Sinitsyn-Nemenman (BSN) phase \cite{Sinitsyn}. 

The method of Ref. \cite{Sagawa} was generalized to quantum systems and applied to studies of the quantum pump \cite{Yuge12,Watanabe,Nakajima}. 
Here we briefly explain the studies of the quantum pump.
At $t=0$ and $t=\tau$, we perform projection measurements of a {\it time-independent} observable $O$ of the baths 
and obtain the outcomes $o(0)$ and $o(\tau)$.
The generating function of $\Dl o=o(\tau)-o(0)$ is 
\bea
Z_{\tau}(\chi)=\int d\Dl o \ P_\tau(\Dl o)e^{i\chi \Dl o}
\eea
where $P_\tau(\Dl o)$ is the probability density distribution of $\Dl o$ and $\chi$ is called the counting field.
To calculate the generating function, the method using the quantum master equation (QME) 
with the counting field (FCS-QME) \cite{FCS-QME} had been proposed.
The solution of the FCS-QME, $\rho^\chi(t)$, provides the generating function as $Z_\tau(\chi)=\tr_S[\rho^\chi(\tau)]$, where
$\tr_S$ denotes the trace of the system. 
The Berry phase \cite{Berry} of the FCS-QME is the BSN phase. 
The average of the difference of the outcomes is given by 
\bea
\bra \Dl o \ket=\int_0^\tau dt \ i^O(t),
\eea
where $i^O(t)$ is the {\it current} of an operator $O$.
If the state of the system at $t=0$ is the instantaneous steady state 
and the modulation of the control parameters is slow, the following relation holds:
\bea
\bra \Dl o \ket=\int_0^\tau dt \ i_\st^O(\al_t)+\int_C d\al^n \ A_n^O(\al), \la{I4}
\eea
where $i_\st^O(\al_t)$ is the instantaneous steady current of $O$.
$C$ is the trajectory from $\al_0$ to $\al_\tau$. 
and $A_n^O(\al)$ is the BSN vector derived from the BSN phase. 
$\al^n$ is $n$-th component of the control parameters and the summation symbol for $n$ is omitted. 
The derived formula of the BSN vector depends on the approximations used for the QME. 
The Born-Markov approximation with or without the rotating wave approximation (RWA) \cite{open} is frequently used. 
The QME in the Born-Markov approximation without RWA sometimes violates the non-negativity of the system reduced density operator.
The  QME of the RWA or the coarse-graining approximation (CGA) \cite{CG,CG13} is the Lindblad type which guarantees the non-negativity \cite{open}.
If $O$ is the total particle number of a bath $b$,   
there are several methods to calculate $A_n^O(\al)$ of \re{I4} \footnote{
For non-interacting system, $A_n^O(\al)$ is calculated from the Brouwer formula\cite{brouwer} 
using the scattering matrix.
Recently, the quantum pump in interacting systems has been actively researched. 
There are three theoretical approaches. 
The first is the Green's function approach, \cite{splettstoesser}. 
The second is the generalized master equation approach \cite{RT09,RT12}. 
The third is the FCS-QME approach. 
Reference\cite{Nakajima} showed the equivalence between the second and the third approaches for all orders of pumping frequency 
(see also \cite{pluecker}).}.

In this paper, we propose to identify 
 \bea
 \dot{\sig}(t)\defe \sum_b \be_b(t)\left[-i^{H_b}(t)-\mu_b(t)\left\{ -i^{N_b}(t)\right\}\right]
\eea
with the {\it average entropy production rate}, 
where $\mu_b$ is the chemical potential of the bath $b$, and $i^{H_b}(t)$ and $i^{N_b}(t)$ are energy and particle currents from the system to the bath $b$, respectively. 
$H_b$ and $N_b$ represent the Hamiltonian and the total particle number of the bath $b$, respectively.
This is a straightforward extention of the entropy production rate argued for an NESS \cite{Utsumi} to a time-dependent system.
Now, the excess entropy is obtained by
\bea
\sig_\RM{ex} = \int_0^\tau dt \ \left[\dot{\sig}(t)- J^\sig_\st(\al_t)\right]=\int_C \ d\al^n \ A_n^\sig(\al), \la{sigex}
\eea
where we used \re{I4} in the second equation with
\bea
A_n^\sig(\al)&\defe& \sum_b \be_b\left[-A_n^{H_b}(\al)-\mu_b\left\{-A_n^{N_b}(\al)\right\}\right] . \la{BSNv}
\eea
Here, 
$A_n^{H_b}(\al)$ and $A_n^{N_b}(\al)$ are the BSN vectors of $H_b$ and $N_b$. 
It should be noted that $\beta_b$ and $\mu_b$ could also be the elements of the set of the control parameters, $\alpha$.
The following expression is the main result of this manuscript,
\bea
A_n^\sig(\al)=-\tr_S\left[\ln {\breve\rho}_0(\al)\f{\p \rho_0(\al)}{\p \al^n} \right]+\mO(\ep^2), \la{Main}
\eea
without any assumption on the time-reversal symmetry.
$\rho_0(\al)$ is the instantaneous steady state of the QME and $\breve{\rho}_0(\al)$ is that of the QME which is given by reversing the sign of the Lamb shift term. 
If the system Hamiltonian is non-degenerate or the Lamb shift term is negligible, we obtain 
\bea 
\sig_\RM{ex} =S_\RM{vN}\left(\rho_0(\al_{\tau})\right)-S_\RM{vN}\left(\rho_0(\al_{0})\right)+\mO(\ep^2 \dl),
\eea
where $S_\RM{vN}(\rho) \defe -\tr_S[\rho \ln\rho]$ is the von Neumann entropy.

The structure of the paper is as follows. 
First, we explain the FCS-QME (\res{sFCS-QME}) and the formula for the excess entropy.
Then we introduce the generalized QME in \res{sGQME}.
In \res{RWA}, we explain the RWA and after this section we focus on the RWA except for \res{Born_Markov}. 
In \res{eq}, the BSN vector $A_n^\sig$ in the equilibrium is discussed. 
In \res{noneq}, the main result of this manuscript, \re{Main}, is derived.
Next we mention the results in the Born-Markov approximation (\res{Born_Markov}). 
In \res{Other_def}, we compare the preceding study on the entropy production in the classical Markov jump process \cite{Komatsu15,Jarzynski} with ours.
In \res{Discussion}, we consider the time-reversal operation. 
At last (\res{Summary}), we summarize this paper. 
In the Appendix \ref{Appendix A}, we derive the formula of the derivative of the von Neumann entropy
and in the Appendix \ref{Appendix B}, we show the details of the derivation
of the relation in a weakly nonequilibrium regime.
In the Appendix \ref{Appendix C}, we explain the definition of entropy production of the Markov jump process 
and the result of Ref. \cite{Komatsu15}.

\section{Quantum master equation and full counting statistics}

\subsection{Full counting statistics-quantum master equation} \la{sFCS-QME}

We consider system $S$ weakly coupled to several baths (although we used the same symbol $S$ as the
entropy in \res{Introduction}, $S$ only means the `system' in the following discussions). 
In order to maintain the system out of equilibrium and in NESS, the system needs to be coupled with more than one bath.
The total Hamiltonian is given by 
\bea
H_\tot(t) = H_S(\al_S(t))+\sum_b \left[H_b+H_{Sb}(\al_{Sb}(t))\right].
\eea
$H_S(\al_S)$ is the system Hamiltonian and $\al_S$ denotes a set of control parameters of the system. 
$H_b$ is the Hamiltonian of the bath $b$. 
$H_{Sb}(\al_{Sb})$ is the coupling Hamiltonian between $S$ and the bath $b$, and $\al_{Sb}$ is corresponding set of control parameters.  
We denote the inverse temperature and the chemical potential of the bath $b$ by $\be_b$ and $\mu_b$ which
can be the control parameters,
and $\al_b$ denotes the set of $\be_b$ and $\be_b \mu_b$.
We symbolize the set of all control parameters $(\al_S$, $\{\al_{Sb}\}_b$, $\al_B\defe \{\al_b \}_b)$ by $\al$. 
While $\al_S$ and $\al_{Sb}$ are dynamical parameters 
like energy levels, tunnel coupling strengths or a magnetic field, 
$\al_B$ are the thermodynamical parameters. 
We denote the set of all the linear operators of $S$ by ${\rm\bm{B}}$. 

Consider slow modulation of the control parameters during $0 \le t \le \tau$. 
At $t=0$ and $t=\tau$, we perform projection measurements of {\it time-independent} observables $\{O_\mu \}$ 
of the baths which commute with each other. 
The index $\mu$ distinguishes the time-independent observables of the baths.
$\{ o_\mu^{(\tau)} \}$ ($\{ o_\mu^{(0)} \}$) denotes the set of outcomes at $t=\tau$ ($t=0$).
The generating function 
\bea
Z_\tau(\{ \chi_{O_\mu} \})=\int \prod_\mu d\Dl o_\mu \ P_\tau(\{\Dl o_\mu \})e^{i\sum_\mu \chi_{O_\mu} \Dl o_\mu}
\eea
is the Fourier transform of the joint probability density distribution $P_\tau(\{\Dl o_\mu \})$ where $\Dl o_\mu \defe o_\mu^{(\tau)}- o_\mu^{(0)} $.
Here, $\chi_{O_\mu}$ is the counting field for $O_\mu$. 
The generating function  is given by 
\bea
Z_\tau(\{ \chi_{O_\mu} \})=\tr_\tot\left[\rho_\tot^\chi(t=\tau)\right]
\eea
 using an operator of the total system $\rho_\tot^\chi(t)$
obeying the modified von Neumann equation \cite{FCS-QME}
\bea 
\f{d}{dt} \rho_\tot^\chi(t) =  -i\left[H_\tot(t), \rho_\tot^\chi(t)\right]_\chi . \la{GLN}
\eea
In this paper, we set $\hbar=1$. 
Here, for two operators $A$ and $B$, $[A,B]_\chi \defe A_{\chi}B-BA_{-\chi}$ and 
\bea
A_\chi \defe e^{i \sum_\mu \chi_{O_\mu} O_\mu/2}Ae^{-i\sum_\mu \chi_{O_\mu} O_\mu/2}. 
\eea
$\chi$ denotes the set of the counting fields $\{ \chi_{O_\mu} \}$. 
The initial condition of $\rho_\tot^\chi(t)$ is given by 
$ \rho_\tot^\chi(0)=\sum_{\{ o_\nu \}} P_{\{ o_\nu \}} \rho_\tot(0) P_{\{ o_\nu \}}$ \cite{FCS-QME}. 
Here, $\rho_\tot(0)$ is the initial state of the total system, $\{ o_\nu \}$ denotes eigenvalues of $\{ O_\nu \}$ and 
$P_{\{ o_\nu \}}$ is a projection operator defined by $O_\mu P_{\{ o_\nu \}}= o_\mu P_{\{ o_\nu \}}$, 
$P_{\{ o_\nu \}}P_{\{ o_\nu^\pr \}}=P_{\{ o_\nu \}} \prod_\mu \dl_{o_\mu,o_\mu^\pr}$, and $P_{\{ o_\nu \}} \dg=P_{\{ o_\nu \}}$. 
We suppose 
\bea
\rho_\tot(0)=\rho(0) \otimes \rho_B(\al_B(0)),
\eea  
where $\rho(0)$ is the initial state of the system and
\bea
\rho_B (\al_B(0)) \defe \bigotimes_{b} \frac{1}{\Xi_b(\al_b(0))}e^{-\beta_b(0) [H_b -\mu_b(0) N_b] }
\eea
with $\Xi_b(\al_b)\defe \tr_b[e^{-\beta_b[ H_b -\mu_b N_b] }]$.  
$\tr_b$ denotes the trace of the bath $b$, and $\tr_B$ denotes the trace over all baths' degrees of freedom. 
Then, we have
\bea
\rho_\tot^\chi(0)=\rho(0) \otimes \sum_{\{ o_\nu \}} P_{\{ o_\nu \}} \rho_B(\al_B(0)) P_{\{ o_\nu \}}.
\eea 
We suppose $[H_b,N_b]=0$. 
If all $O_\mu$ commute with $H_b$ and $N_b$, $P_{\{ o_\nu \}}$ commutes with $\rho_B(\al_B(0))$ and 
$\rho_\tot^\chi(0)=\rho(0) \otimes \rho_B(\al_B(0))$ holds because $\sum_{\{ o_\nu \}} P_{\{ o_\nu \}}=1$.

We defined $\rho^\chi(t)\defe \tr_B[\rho_\tot^\chi(t)]$ and the generating function is calculated with 
\bea
Z_\tau(\{ \chi_{O_\mu} \})=\tr_S[\rho^\chi(t=\tau)].
\eea
We assume 
\bea
\rho_\tot^\chi(t)\approx \rho^\chi(t) \otimes \rho_B(\al_B(t))\ \ \ (0<t\le \tau)
\eea
where 
\bea
\rho_B (\al_B(t)) \defe \bigotimes_{b} \frac{1}{\Xi_b(\al_b(t)) }e^{-\beta_b(t) [H_b -\mu_b(t) N_b] }.
\eea
The FCS-QME \cite{FCS-QME} is 
\bea
\f{d \rho^\chi(t)}{dt}=\hat{K}^\chi(\al_t)\rho^\chi(t), \la{FCS-QME}
\eea
and the initial condition is $\rho^\chi(0)=\rho(0)$. 
Here $\hat{K}^\chi(\al_t)$ is the Liouvillian modified by $\chi$.
The Liouvillian is given by 
\bea 
\hat{K}^\chi(\al)\bu=-i[H_S(\al_S),\bu]+\sum_b \mL_b^\chi(\al) \bu. 
\eea
Here and in the following, $\bu \in {\rm\bm{B}}$. 
$\mL_b^\chi(\al) $ describes the coupling effects between $S$ and the bath $b$ and
depends on used approximations, for instance, the Born-Markov approximation without or within the RWA and the CGA. 
Generally, $\mL_b^\chi(\al) $ has the form: 
\bea
\mL_b^\chi(\al)  \bu=\sum_a c_{ba}^\chi(\al) A_a \bu B_a , \la{mL_b}
\eea
where $A_a$ and $B_a$ belong to ${\rm\bm{B}}$ and depend on $\al_S$, 
and $c_{ba}^\chi(\al)$ is a complex number which depends on $\al_S$, $\al_{Sb}$ and $\al_b$.  
If and only if $A_a, B_a \ne 1$, $c_{ba}^\chi(\al)$ depends on $\chi$. 
After \res{RWA} we choose the Born-Markov approximation within RWA; 
however, in this subsection we assume only Markov property (i.e., $\hat{K}^\chi$ just depends on $\al_t$).
Explicit expression of (\ref{mL_b}) will be given in \res{RWA} .
At $\chi=0$, the FCS-QME becomes the quantum master equation (QME)
\bea
 \f{d\rho(t)}{dt} = \hat{K}(\al_t)\rho(t). \la{QME}
\eea
$\hat{K}(\al_t)$ equals $\hat{K}^\chi(\al_t)$ at $\chi=0$. In the following, a symbol $X$ without $\chi$ denotes $X^\chi \vert_{\chi=0}$.

In the Liouville space \cite{Nakajima,FCS-QME}, the left and right eigenvalue equations of the Liouvillian are
\bea
\hat{K}^\chi(\al)\dke{\rho_n^\chi(\al)}\aeq \lm_n^\chi(\al)\dke{\rho_n^\chi(\al)} \la{rig} ,\\
\dbr{l_n^\chi(\al)}\hat{K}^\chi(\al)\aeq \lm_n^\chi(\al)\dbr{l_n^\chi(\al)} \la{left}.
\eea
In the Liouville space, $A \in {\rm\bm{B}}$ is described by $\dke{A}$. 
The inner produce is defined by $\dbr{A}B\dket= \tr_S(A\dg B)$ ($A,B \in {\rm\bm{B}}$). 
In particular, $\dbr{1}A\dket= \tr_S A$ holds. 
A superoperator which operates to a liner operator of the system becomes an operator of the Liouville space. 
The left eigenvectors $l_n^\chi(\al)$ and the right eigenvectors $\rho_m^\chi(\al)$ satisfy $\dbra l_n^\chi(\al) \dke{\rho_m^\chi(\al)}=\dl_{nm}$. 
The mode which has the eigenvalue $\lm_n^\chi(\al)$ with the maximum real part is assigned by the label $n=0$.  
Because the conservation of the probability, $\dbra 1 \dke{\rho(t)}=1$, and using (\ref{QME}), the relation
\bea
 \f{d}{dt}\dbra 1 \dke{\rho(t)}=\dbr{1}\hat{K}(\al_t)\dke{\rho(t)}=0
 \eea
leads
$\dbr{1}\hat{K}(\al)=0$, in the limit $\chi \to 0$, $\lm_0^\chi(\al)$ becomes $0$ and $\dbr{l_0^\chi(\al)}$ becomes $\dbr{1}$
(i.e., $l_0(\al)$ is an identity operator). 
In addition, the special state $\dke{\rho_0(\al)}$ determined by $\hat{K}(\al)\dke{\rho_0(\al)}=0$ 
represents the {\it instantaneous steady state}.

The formal solution of the FCS-QME \re{FCS-QME} is
\bea
\dke{\rho^\chi(t)} = {\rm{T}} \exp \left[\int_0^t ds \ \hat{K}^\chi(\al_s) \right]\dke{\rho(0)} ,
\eea
where ${\rm{T}}$ denotes the time-ordering operation. Using this, 
we obtain the averages of $\Dl o_\mu$ at time $t$
\bea
\bra \Dl o_\mu \ket_t \aeq \f{\partial }{\partial(i\chi_{O_\mu}) } \dbra 1 \dke{\rho^\chi(t)}   \Big \vert _{\chi=0} \no\\
\aeq  \int_0^t du \ \dbr{1}\hat{K}^{O_\mu}(\al_u)\dke{\rho(u)} \e  \int_0^t du \  i^{O_\mu} (u), \la{wat}
\eea
where 
\bea
X^{O_\mu}(\al)\defe \f{\partial X^\chi(\al) }{\partial (i\chi_{O_\mu})} \big \vert_{\chi=0},
\eea
 when $X$ is an (super)operator or a c-number
and $i^{O_\mu}(u)$ is the {\it current} of $O_\mu$ at time $u$.
Here, we used $\dbr{1}\hat{K}(\al)=0$.
Moreover, using $\dbr{l_0(\al)}=\dbr{1}$, $\lm_0(\al)=0$ and \re{left}, we obtain
\bea
\dbr{1}\hat{K}^{O_\mu}(\al)
= \lm_0^{O_\mu} (\al) \dbr{1} - \dbr{l_0^{O_\mu} (\al)}\hat{K}(\al). \la{wata}
\eea
Here, $\dbr{l_0^{O_\mu} (\al)}$ is defined by $\f{\partial \dbr{l_0^\chi(\al)} }{\partial (i\chi_{O_\mu})} \big \vert_{\chi=0}$, 
then $l_0^{O_\mu}=-\f{\partial l_0^\chi(\al) }{\partial (i\chi_{O_\mu})} \big \vert_{\chi=0}$ holds.
Now, the current $i^{O_\mu} (t)$ is given by \cite{Watanabe}
\bea
i^{O_\mu} (t) \aeq \dbr{1}\hat{K}^{O_\mu}(\al_t)\dke{\rho(t)} \no\\
\aeq \lm_0^{O_\mu} (\al_t)\dbra 1 \dke{\rho(t)} - \dbr{l_0^{O_\mu} (\al_t)}\hat{K}(\al_t) \dke{\rho(t)}  \no\\
\aeq \lm_0^{O_\mu} (\al_t)- \dbr{l_0^{O_\mu} (\al_t)}\f{d}{dt}\dke{\rho(t)}. \la{watan}
\eea
The current can also be written as
\bea
i^{O_\mu}(t) 
= \dbr{1} W^{O_\mu} (\al_t) \dke{\rho(t)} , \la{I_mu}
\eea
where $W^{O_\mu}(\al)$ is the current operator defined by 
\bea
\dbr{1}W^{O_\mu}(\al) = \dbr{1}\hat{K}^{O_\mu} (\al) \la{defW},
\eea
i.e., $\tr_S[W^{O_\mu}(\al) \bu] =\tr_S[\hat{K}^{O_\mu}(\al)\bu]$ for any $\bu \in {\rm\bm{B}}$. 
Therefore, using \re{mL_b}, the current operator is given by 
\bea
 W^{O_\mu}(\al)=\sum_{b,a} c_{ba}^{O_\mu}(\al) B_a A_a . \la{W^O}
\eea
Using \re{wata}, the {\it instantaneous steady current} is given by 
\bea
\dbr{1}W^{O_\mu} (\al) \dke{\rho_0(\al)} = \lm_0^{O_\mu}(\al)\e i^{O_\mu}_\st(\al) . \la{touka_sc}
\eea
In the following, we suppose 
that the state of the system at $t=0$ is the instantaneous steady state,
$\rho(0)=\rho_0(\al_0)$. 
Then, $\rho(t)=\rho_0(\al_t)+\mO(\om/\Ga)$ holds \cite{Nakajima} where 
$\om=2\pi/\tau$ and  $\Ga=\min_{n \ne 0}\{-\RM{Re}(\lm_n)\}$. 
In $\om \ll \Ga$ limit, we obtain  
\bea
i^{O_\mu}(t)  =  i^{O_\mu}_\st(\al_t)-\dbr{l_0^{O_\mu} (\al_t)}\f{d}{dt}\dke{\rho_0(\al_t)}+\mO\big(\f{\om^2}{\Ga}\big), \la{i^O}
\eea
which leads to
\bea
\bra \Dl o_\mu \ket_\tau = \int_0^\tau dt \ i^{O_\mu}_\st(\al_t)+\int_C d\al^n \ A_n^{O_\mu}(\al) +\mO\big(\f{\om}{\Ga}\big), \la{Dl o}
\eea
where in the second term, the summation symbol $\sum_{n}$ is omitted. 
Here, $\al^n$ is the $n$-th component of the control parameters,  $C$ is the trajectory from $\al_0$ to $\al_\tau$, and
\bea
A_n^{O_\mu}(\al) \defe -\dbr{l_0^{O_\mu} (\al)}\f{\p }{\p \al^n}\dke{\rho_0(\al)}, \la{BSN}
\eea
is the BSN vector.
The BSN vector is also given by \cite{Nakajima}
\bea
A_n^{O_\mu}(\al) =\dbr{1}W^{O_\mu}(\al)\mR(\al)\f{\p }{\p \al^n}\dke{\rho_0(\al)}, \la{BSN_Nakajima}
\eea 
where  $\mR(\al)$ is the pseudoinverse of the Liouvillian defined by
\bea
\mR(\al)\hat{K}(\al)=1-\dke{\rho_0(\al)}\dbr{1}. \la{defR}
\eea
The expression of \re{Dl o} was originally derived like the following.
The formal solution of the FCS-QME is expanded as
\bea
\dke{\rho^\chi(t)} = \sum_n c_n^\chi(t)e^{\int_0^t ds \ \lm_n^\chi(\al_s)}\dke{\rho_n^\chi(\al_t)} \la{exp}.
\eea
Because $e^{\int_0^t ds \ \lm_n^\chi(\al_s)}$ $(n \ne 0)$ exponentially damps as a function of time, 
only $n=0$ term remains if $\Ga \tau \gg 1$. 
Solving the time evolution equation of $c_0^\chi(t)$ in $\om \ll \Ga$ limit, we obtain
\bea
c_0^\chi(\tau) = c_0^\chi(0)\exp \left[-\int_0^\tau dt \  \dbr{l_0^\chi(\al_t)} \f{d }{dt}\dke{\rho_0^\chi(\al_t)} \right].
\eea
Here, the argument of the exponential function is called the BSN phase. 
Substituting this  expression and $c_0^\chi(0)=\dbra l_0^\chi(\al_0)\dke{\rho_0(\al_0)}$ into \re{exp}, 
we obtain the expression of $\rho^\chi(\tau)$ which provides \re{Dl o}. 
However, when we consider only the average of $\Dl o_\mu$, the BSN phase is not essential. 
All informations of the counting fields up to the first order are included in $W^{O_\mu}$ \footnote{
In the research of adiabatic pumping, the expression of \re{Dl o} is essential. 
In Refs.\cite{Yuge12,Watanabe,Nakajima}, \re{Dl o} with \re{BSN} was used to study the quantum pump.
On the other hand, in Ref. \cite{RT12}, \re{Dl o} was derived using the generalized master equation \cite{RT09} and without using the FCS. 
In Ref. \cite{RT12}, $A_n^{O_\mu}(\al)$ was described by the quantity corresponding to the current operator and the pseudoinverse of the Liouvillian, 
as shown in \re{BSN_Nakajima}.    
Reference\cite{Nakajima} showed the equivalence between the FCS-QME approach and the generalized master equation approach for all orders of pumping frequency.}

As discussed in \res{Introduction}, we propose to identify the average entropy production rate with
\bea
\dot{\sig}(t)\defe \sum_b \be_b(t)\left[-i^{H_b}(t)-\mu_b(t)\left\{ -i^{N_b}(t)\right\}\right] .
\eea
This is given by $\dot{\sig}(t)=\tr_S[W^\sig (\al_t)\rho(t)]$ with 
\bea
W^\sig (\al) \defe \sum_b \be_b[-W^{H_b}(\al)-\mu_b\{ -W^{N_b}(\al)\}]. \la{def_W^sig}
\eea
The average entropy production is given by
\bea
\sig \aeqd \int_0^\tau dt\ \dot{\sig}(t) \no\\
\aeq \int_0^\tau dt \ J^\sig_\st(\al_t)+\int_C \ d\al^n \ A_n^\sig(\al) +\mO\big(\f{\om}{\Ga}\big)\la{def_sig},
\eea
where 
\bea
J^\sig_\st(\al)\defe \sum_b \be_b[-i_\st^{H_b}(\al)-\mu_b\{-i_\st^{N_b}(\al)\}]
\eea
 and 
$A_n^\sig(\al)$ is defined in (\ref{BSNv}).
Here, we used \re{i^O} for $O_\mu=H_b,N_b$. 
The excess entropy production is defined as (\ref{sigex}) by 
\bea
\sig_\RM{ex} \defe \int_C \ d\al^n \ A_n^\sig(\al)+\mO\big(\f{\om}{\Ga}\big) .
\eea

\subsection{Generalized quantum master equation for entropy production} \la{sGQME}

We consider a kind of generalized quantum master equation (GQME) 
\bea
\f{d}{dt}\rho^\lm(t) = \mK^\lm(\al_t)\rho^\lm(t) ,\la{GQME}
\eea
with the initial condition $\rho^\lm(0)=\rho(0)$. 
Here, $\lm$ is a single real parameter.
We suppose that the Liouvillian is given by 
\bea
\mK^\lm(\al)\bu=-i[H_S(\al_S),\bu]+\sum_b \mL_b^\lm(\al) \bu
\eea
 with $\mL_b^\lm (\al)\bu=\sum_a c_{ba}^\lm(\al) A_a \bu B_a$ 
and $c_{ba}^\lm(\al)\bv{\lm=0}=c_{ba}$. 
While $c_{ba}^\chi(\al)$ of \re{mL_b} depends on $\chi$ if and only if $A_a, B_a \ne 1$, 
$c_{ba}^\lm(\al)$ can depend on $\lm$ for all $a$. 
We suppose that the solution of \re{GQME} satisfies
\bea
\tr_S[\rho^\pr(\tau)] = \sig ,\la{G_def} 
\eea
where $X^\pr \defe \f{\p X^\lm}{\p (i\lm)}\Bv{\lm=0}$. 
This condition is equivalent to 
\bea
\dbr{1}\mK^\pr(\al) = \dbr{1}W^\sig(\al). \la{G_con}
\eea
Let's consider 
\bea 
\dbr{l_0^\lm(\al)}\mK^\lm(\al)=\lm_0^\lm(\al)\dbr{l_0^\lm(\al)} \la{G_l_0} ,
\eea
corresponding to \re{left} for $n=0$.
Similar to \re{touka_sc} and \re{BSN}, 
\bea
\lm_0^\pr(\al) \aeq \dbr{1}W^\sig(\al)\dke{\rho_0(\al)}=J_\st^\sig(\al), \la{lm_0^pr}
\eea
and 
\bea
A_n^\sig(\al) \aeq -\dbr{l_0^\pr (\al)}\f{\p }{\p \al^n}\dke{\rho_0(\al)}=\dbr{1}W^\sig(\al)\mR(\al)\f{\p }{\p \al^n}\dke{\rho_0(\al)}, \la{A_n^sig_F}
\eea
hold. 
Although $\lm_0^\lm(\al)$ and $l_0^\lm(\al)$ depend on the choice of $\mK^\lm(\al)$, $\lm_0^\pr(\al)$ and $A_n^\sig(\al)$ do not depend, 
as can be seen in the right-hand side (RHS) of the \re{lm_0^pr} and \re{A_n^sig_F}.
The LHS of \re{G_con} is given by
\bea
\dbr{1}\mK^\pr(\al) = \dbr{1} \sum_{b,a} c_{ba}^\pr(\al) B_a A_a .
\eea
Using this and \re{W^O}, \re{G_con} becomes
\bea
\sum_{b,a} c_{ba}^\pr(\al) B_a A_a = \sum_{b,a} \left[-\be_b c_{ba}^{H_b}(\al)+\be_b \mu_b c_{ba}^{N_b}(\al) \right] B_a A_a .
\eea
Infinite solutions of this equation exist. 
One choice of $\mK^\lm(\al)$ satisfying this relation is $\hat{K}^\chi(\al)$ in the limit of $\chi_{H_b} \to -\be_b \lm$ and $\chi_{N_b} \to \be_b\mu_b \lm$.

While we can calculate the average of the entropy production as shown in \res{sFCS-QME} and in this subsection, 
our formalism is not compatible to discuss the higher moments of the entropy production.  
``Higher moments'' $\f{\p^n}{\p (i\lm)^n}\tr_S[\rho^\lm(\tau)]\bv{\lm=0}$ ($n=2,3,\cdots$) depend on the choice of $\mK^\lm(\al)$ and 
currently there seems no physical guiding principle to determine an adequate $\mK^\lm(\al)$.
Although \re{Dl o} is the average of the difference between outcomes at $t=\tau$ and $t=0$ of $O_\mu$, 
there is no bath's operator corresponding to $\sig$ if $\al_B$ are modulated.
In contrast, the higher moments of the entropy production could be considered for the classical Markov jump process.
In Appendix \ref{Appendix C}, we review the entropy production of the Markov jump process \cite{Komatsu15,Jarzynski}, and in \res{Other_def}, we compare that and \re{def_sig}. 

\subsection{Rotating wave approximation} \la{RWA}
 
In this subsection, we introduce the FCS-QME within RWA. 
First, we introduce the CGA. 
An operator in the interaction picture corresponding to $A(t)$ is defined by $A^I(t)=U_0\dg(t)A(t)U_0(t)$ with 
\bea
\f{dU_0(t)}{dt}=-i\left[H_S(\al_S(t))+\sum_{b}H_b\right]U_0(t) 
\eea
and $U_0(0)=1$.
The system reduced density operator in the interaction picture is given by $\rho^{I,\chi}(t)=\tr_B[\rho_\tot^{I,\chi}(t)]$ 
where $\rho_\tot^{I,\chi}(t)=U_0(t)\rho_\tot^\chi(t)U_0\dg(t)$. 
$\rho_\tot^{I,\chi}(t)$ is governed by
\bea
\f{d\rho_\tot^{I,\chi}(t)}{dt} = -i[H_{\rm{int}}^I(t),\rho_\tot^{I,\chi}(t)]_\chi, \la{vN,I}
\eea
with $H_{\rm{int}}=\sum_{b}H_{Sb}$. 
Up to the second order perturbation in $H_{\rm{int}}$, we obtain
\bea
\rho^{I,\chi}(t+\tau_{\rm{CG}}) 
\aeq \rho^{I,\chi}(t)\no \\
&-& \int_t^{t+\tau_{\rm{CG}}} du \int_t^{u} ds \ 
\tr_B \left\{ [ H_{\rm{int}}^I(u), [H_{\rm{int}}^I(s), \rho^{I,\chi}(t) \otimes \rho_B (\al_B(t)) ]_\chi ]_\chi \right \} \no\\
\aeqe \rho^{I,\chi}(t) +\tau_{\rm{CG}}\hat{L}_{\tau_{\rm{CG}}}^\chi(t)\rho^{I,\chi}(t) , \la{FCS-QME_I_CG}
\eea
using the large-reservoir approximation:
\bea
\rho_\tot^{I,\chi}(t) \approx \rho^{I,\chi}(t) \otimes \rho_B (\al_B(t))
\eea
and 
$\tr_B[H_{\rm{int}}^I(u)\rho_B (\al_B(t))]=0$. 
The arbitrary parameter $\tau_{\rm{CG}}$ $(>0)$ is called the coarse-graining time. 
The CGA \cite{CG,CG13} is defined by 
\bea
\f{d}{dt}\rho^{I,\chi}(t)=\hat{L}_{\tau_{\rm{CG}}}^\chi(t)\rho^{I,\chi}(t). \la{def,CGA}
\eea
At $\chi=0$, this is Lindblad type.
If $\tau \gg \tau_{\rm{CG}}$, 
the superoperator $\hat{L}_{\tau_{\rm{CG}}}^\chi(t)$ is described as a function of the set of control parameters at time $t$. 
In this paper, we suppose $\tau \gg \tau_{\rm{CG}}$. 
Moreover, $\tau_{\rm{CG}}$ should be much shorter than the relaxation time of the system, $\tau_S\sim \f{1}{\Ga}$. 
On the other hand, $\tau_S \ll \tau $ should hold for the adiabatic modulation. 
Hence $\tau_{\rm{CG}} \ll \f{1}{\Ga} \ll \tau$ should hold.  
By the way, the Born-Markov approximation is given by 
\bea
\f{d\rho^{I,\chi}(t)}{dt} = -\int_0^\infty ds \ \tr_B\Big\{[H_{\rm{int}}^I(t),[H_{\rm{int}}^I(t-s),\rho^{I,\chi}(t)\otimes \rho_B (\al_B(t))]_\chi]_\chi \Big\}.\la{BM}
\eea 

Now we suppose
\bea
H_{Sb}(\al_{Sb})= \sum_{k,\al}V_{bk,\al }(\al_{Sb})a_{\al}\dg c_{bk}+\hc , \la{H_Sb}
\eea
where $a_{\al}$ and $c_{bk}$ are single-particle annihilation operators of the system and of the bath $b$. 
Although we have used indeces $\al$ or $\beta$ to distinguish the system operators, this may not
confuse the readers with the set of control parameters or the inverse temperature since
they only appear as a subscript of the operator $a$ (or $a^\dagger$) and the parameters like $V_{bk,\al}, \Phi_{b,\al\beta}^{\pm}, \Psi_{b,\al\beta}^{\pm}$
or under the summation symbol.
The eigenoperator defined by  
\bea
a_{\al}(\om) = \sum_{n,r,m,s} \dl_{\om_{mn},\om} \ke{E_n,r}\br{E_n,r} a_{\al} \ke{E_m,s} \br{E_m,s}. \la{Def_a_om}
\eea
is useful to describe the FCS-QME. 
Here, $ \om_{mn}\defe E_m-E_n$, 
\bea
H_S\ke{E_n,r}=E_n\ke{E_n,r}
\eea
 and $r$ denotes the label of the degeneracy.
$\om$ is one of the elements of 
\bea
\mathcal{W}=\{ \om_{mn} \vert \ {  \br{E_n,r} a_{\al} \ke{E_m,s} \ne 0 \hs{2mm} }^\exists \al \} . 
\eea
$a_{\al}(\om)$ and $\om$ depend on $\al_S$. 
$\sum_\om a_{\al}(\om)=a_{\al}$,
\bea
[H_S,a_{\al}(\om)]=-\om a_{\al}(\om)\ \ \mbox{and}\ \  [N_S,a_{\al}(\om)]=-a_{\al}(\om)
\eea 
hold.
Here, $N_S$ is total number operator of the system. 
We suppose $[N_S,H_S]=0$.
In the CGA or Born-Markov approximation, the FCS-QME is described by $a_{\al}(\om)$ and $[a_{\al}(\om^\pr)]\dg$ $(\om,\om^\pr \in \mathcal{W})$. 
If $H_S$ is time dependent, the generalization of usual RWA \cite{open} with static $H_S$ is unclear. 
In this paper, the RWA is defined as the limit $\tau_\RM{CG} \to \infty$ ($\tau_\RM{CG} \cdot \min_{\om \ne \om^\pr}\abs{\om-\om^\pr} \gg 1$) of the CGA.  
If $H_S$ is time independent, this RWA is equivalent to usual RWA. 

In the following, except for \res{Born_Markov}, we consider the RWA. 
Then, $\mL_b^\chi(\al)$ is generally given by (for the details of the derivation, please refer \cite{Nakajima,FCS-QME})
\bea
\mL_b^\chi(\al) \bu =\Pi_b^\chi(\al) \bu -i[h_b(\al),\bu], \la{def_Pi_b}
\eea
where $h_b(\al)$ is a Hermitian operator describing the {\it Lamb shift}.
$h_b(\al)$ commutes with $H_S(\al_S)$ for general model and with $N_S$ for the model \re{H_Sb}.
$H_\RM{L}(\al)\defe \sum_b h_b(\al)$ is called {\it Lamb shift Hamiltonian}.
The superoperator $\Pi_b^\chi(\al)$ represents the dissipation.

Here, we suppose the free Hamiltonian of the bath $b$:
\bea
H_b = \sum_{k} \ep_{bk}c_{bk} \dg c_{bk},
\eea
and $\{O_\mu\}=\{N_b,H_b\}_b$. 
 $\Pi_b^\chi(\al)$ in \re{def_Pi_b} is given by
\bea
\Pi_b^\chi(\al) \bu 
\aeq    \sum_{\om }  \sum_{\al,\be} \Big[
 \Phi^{-,\chi}_{b,\al  \be }(\om) a_{\be }(\om) \bu [a_{\al} (\om)]\dg 
-\half \Phi^-_{b,\al \be}(\om) \bu [a_{\al}(\om)] \dg a_{\be}(\om)\no\\
 \no\\
&&-\half \Phi^-_{b,\al  \be }(\om) [ a_{\al }  (\om)]\dg a_{\be }(\om) \bu+\Phi^{+,\chi}_{b,\al  \be }(\om) [a_{\be }  (\om)]\dg \bu a_{\al } (\om) \no\\
&&-\half \Phi^+_{b,\al  \be }(\om) \bu a_{\al }(\om) [a_{\be }  (\om)]\dg 
-\half \Phi^+_{b,\al  \be }(\om)  a_{\al }  (\om)[a_{\be }  (\om)]\dg \bu  \Big] ,
\eea
where 
\bea
\Phi_{b,\al \be }^{-,\chi}(\Om) \aeq 2\pi\sum_{k}  V_{bk,\al } V_{bk,\be }^\ast
F_b^-(\ep_{bk}) e^{i\chi_{N_b} } e^{i\chi_{H_b} \ep_{bk}}  \dl(\ep_{bk}-\Om)\no\\
\aeq  e^{i\chi_{N_b} +i\chi_{H_b}\Om}\Phi_{b,\al \be }^{-}(\Om) , \la{phi-}\\
\Phi_{b,\al \be}^{+,\chi}(\Om) \aeq 2\pi\sum_{k} V_{bk,\al }^\ast V_{bk, \be }
F_b^+(\ep_{bk}) e^{-i\chi_{N_b} } e^{-i\chi_{H_b} \ep_{bk}}  \dl(\ep_{bk}-\Om)\no\\
\aeq e^{-i\chi_{N_b} -i\chi_{H_b}\Om}\Phi_{b,\al \be }^{+}(\Om). \la{phi+}
\eea
Here, $\chi_{N_b}$ and $\chi_{H_b}$ are the counting fields for $N_b$ and $H_b$. 
If the baths are fermions, 
\bea
F_b^+(\ep)=f_b(\ep)\defe\frac{1}{e^{\be_b(\ep-\mu_b)}+1}
\eea
 and $F_b^-(\ep)=1-f_b(\ep)$. 
If  the baths are bosons,
\bea
F_b^+(\ep)=n_b(\ep)\defe\frac{1}{e^{\be_b(\ep-\mu_b)}-1}
\eea
 and $F_b^-(\ep)=1+n_b(\ep)$.
The Lamb shift is given by
\bea
h_b(\al) \aeq \sum_\om \sum_{\al,\be} \Big( -\half \Psi^-_{b,\al  \be }(\om) [ a_{\al }  (\om)]\dg a_{\be }(\om) 
+\half \Psi^+_{b,\al \be}(\om)  a_{\al }  (\om)[a_{\be }  (\om)]\dg \Big),\no\\
\eea
where 
\bea
\Psi_{b,\al \be }^-(\Om) \aeq 2\sum_{k}  V_{bk,\al } V_{bk,\be }^\ast  F_b^-(\ep_{bk})  {\rm{P}}\f{1 }{\ep_{bk}-\Om} , \\
\Psi_{b,\al \be }^+(\Om) \aeq 2\sum_{k} V_{bk,\al }^\ast V_{bk, \be }F_b^+(\ep_{bk})  {\rm{P}}\f{1 }{\ep_{bk}-\Om} .
\eea
Here, $\rm{P}$ denotes the Cauchy principal value.
$\Phi_{b,\al \be }^{\pm}(\Om)$ satisfy 
\bea
[\Phi_{b,\al \be }^{\pm}(\Om)]^\ast \aeq \Phi_{b, \be \al}^{\pm}(\Om) ,\\
\Phi_{b,\al \be }^+(\Om) \aeq e^{-\be_b(\Om-\mu_b)}\Phi_{b, \be \al}^-(\Om) .\la{KMS}
\eea
The latter is the Kubo-Martin-Schwinger (KMS) condition.

We introduce projection superoperators $\mP(\al_S)$ and $\mQ(\al_S)$ by 
\bea
\mP(\al_S) \ke{E_n,r}\br{E_m,s}=\dl_{E_n,E_m}\ke{E_n,r}\br{E_m,s},
\eea
and $\mQ(\al_S)=1-\mP(\al_S)$. We define sets of operators ${\rm\bm{B}_P}\defe\{X\in {\rm\bm{B}} \vert \mP X=X  \}$
 and ${\rm\bm{B}_Q}\defe\{X\in {\rm\bm{B}} \vert \mQ X=X  \}$.
$\hat{K}^\chi\mP \bu \in {\rm\bm{B}_P}$ holds. 
Then, $\hat{K}^\chi\mQ \bu \in {\rm\bm{B}_Q}$ and
\bea
\mQ\hat{K}^\chi \mP=0=\mP\hat{K}^\chi \mQ ,\la{bunri}
\eea
hold. 
This implies that the right eigenvalue equations \re{rig} are decomposed into two closed systems of equations 
for $\mP\rho_n^\chi$ and for $\mQ\rho_n^\chi$.
Thus, $\rho_n^\chi$ is an element of ${\rm\bm{B}_P}$ or ${\rm\bm{B}_Q}$.
In particular, $\rho_0^\chi \in {\rm\bm{B}_P}$.
Then, the matrix representation of $\rho_0(\al)$ by $\ke{E_n,r}$ is block diagonalized.
This implies
\bea
[H_S(\al_S),\rho_0(\al)]= 0 . \la{H_S,rho_0}
\eea

The particle and energy current operators from the system into bath $b$, $w^{N_b}(\al)$ and $w^{H_b}(\al)$, 
are usually defined by 
\bea
w^{X_b}(\al) \defe -[\mL_b\dg(\al) X_S]\dg=-\mL_b\dg(\al) X_S \ \ \ (X=N,H). \la{def_w}
\eea
For a superoperator $\mJ$, $\mJ\dg$ is defined by $\dbr{\mJ\dg X}Y \dket=\dbra X \dke{\mJ Y} $ ($X,Y \in {\rm\bm{B}}$).
$\mL_b\dg(\al) \bu =\sum_a c_{ba}^\ast(\al) A_a\dg \bu B_a\dg$ holds. 
$w^{X_b}(\al)$ is a Hermitian operator and is given by 
\bea
w^{X_b}(\al) =- \sum_a c_{ba}(\al) B_a X_S A_a \ \ \ (X=N,H) . \la{W_A_B}
\eea
For the Born-Markov approximation and the CGA, $w^{N_b}(\al)=W^{N_b}$, while
$w^{H_b}(\al)\ne W^{H_b}(\al)$. 
For RWA, 
\bea
w^{N_b}(\al)\aeq W^{N_b}(\al)\no\\
\aeq \sum_{\om}\sum_{\al,\beta}  \Big\{   \Phi_{b,\al\beta }^-(\om) [a_{\al }(\om)] \dg a_{\beta }(\om) 
-   \Phi_{b,\al \beta}^+(\om)a_{\al }  (\om)  [a_{\beta } (\om)]\dg  \Big\} ,\la{w=W} \\
w^{H_b}(\al)\aeq W^{H_b}(\al)\no\\
\aeq \sum_{\om}\sum_{\al,\beta}  \Big\{  \om \Phi_{b,\al \beta }^-(\om) [a_{\al }(\om)] \dg a_{\beta }(\om) 
-  \om \Phi_{b,\al \beta }^+(\om)a_{\al }  (\om)  [a_{\beta } (\om)]\dg  \Big\}, \la{w=W2}
\eea
hold. 
Therefore, \re{def_W^sig} and \re{def_w} imply that
$W^\sig(\al)$ is given by 
\bea
W^\sig(\al) = \sum_b \mL_b\dg(\al) (\be_b H_S-\be_b \mu_b N_S)=\sum_b \Pi_b\dg(\al) (\be_b H_S-\be_b \mu_b N_S) .\la{W^sig}
\eea

\section{Geometrical expression of excess entropy production} 

\subsection{Equilibrium state} \la{eq}

In this subsection, we consider equilibrium state $\be_b=\be$ and $\mu_b=\mu$, and $\al$ denotes the set of $(\al_S$, $\{\al_{Sb}\}_b$, $\be$, $\be \mu)$.
We show that $A_n^\sig(\al)$ is a total derivative of the von Neumann entropy of the instantaneous steady state.
Differentiating \re{G_l_0} by $i\lm$ and setting $\lm=0$, we obtain
\bea
\dbr{l_0^\pr(\al)}\hat{K}(\al)+\dbr{1}\mK^\pr(\al) = \lm_0^\pr(\al)\dbr{1}. \la{ee}
\eea
In the RHS, $\lm_0^\pr(\al)=J_\st^\sig(\al)=0$ holds. 
The second term of the LHS is $\dbr{1}W^\sig(\al)$.
\re{W^sig} leads 
\bea
W^\sig(\al) = \be \sum_b  \mL_b\dg(\al)[ H_S-\mu N_S]=\be \hat{K}\dg(\al)[H_S-\mu N_S],
\eea
i.e., 
\bea
\dbr{\be[H_S-\mu N_S]} \hat{K}(\al) = \dbr{1}W^\sig(\al).
\eea
Then, \re{ee} leads 
\bea
\big[\dbr{l_0^\pr(\al)}+\dbr{\be[H_S-\mu N_S]} \big] \hat{K}(\al) = 0.
\eea
This implies 
\bea
\dbr{l_0^\pr(\al)} = -\dbr{\be[H_S-\mu N_S]} +c(\al)\dbr{1} ,
\eea
i.e., $\{l_0^\pr(\al)\}\dg = -\be[H_S-\mu N_S]+c(\al)$ where $c(\al)$ is an unimportant complex number. 
The equilibrium state, $\rho_0(\al)$, is given by
\bea
\rho_0(\al) = \rho_\RM{gc}(\al_S;\be,\be \mu) \defe \f{e^{-\be(H_S(\al_S)- \mu N_S)}}{\Xi(\al_S;\be,\be \mu)} ,
\eea
with $\Xi(\al_S;\be,\be \mu)\defe\tr_S[e^{-\be(H_S(\al_S)-\mu N_S)}]$.
Then, 
\bea
\{l_0^\pr(\al)\}\dg = \ln\rho_\RM{gc}(\al_S;\be,\be \mu) +c^\pr(\al) 1
\eea
with $c^\pr(\al)=c(\al)+\ln\Xi(\al_S;\be,\be \mu)$, holds. 
Substituting this equation into \re{A_n^sig_F}, we obtain 
\bea
A_n^\sig(\al) = \f{\partial }{\partial \al^n} S_\RM{vN}(\rho_\RM{gc}(\al_S;\be,\be \mu)) ,
\eea
where we used \re{dS} in the Appendix \ref{Appendix A}.

\subsection{Weakly nonequilibrium regime} \la{noneq}
In this subsection, we study the BSN vector and the excess entropy production in a weakly nonequilibrium condition.
We introduce parameters characterizing the degree of nonequilibrium:
\bea
\ep_{1,b}\defe \be_b -\overline{\be},\ \ep_{2,b}\defe \be_b\mu_b -\overline{\be\mu} ,\ \ep \defe \max_{b}\big\{\f{\abs{\ep_{1,b}}}{\overline{\be}},
\f{\abs{\ep_{2,b}}}{\abs{\overline{\be\mu}}} \big\}, 
\eea
where $\overline{\be}$ and $\overline{\be \mu}$ are the reference values, 
which satisfy $\min_b \be_b \le \overline{\be} \le \max_b \be_b $ and $\min_b \be_b\mu_b \le \overline{\be\mu} \le \max_b\mu_b \be_b $. 
$\ep$ is a measure of degree of nonequilibrium.
We consider $\ep \ll 1$ regime. 
Now, we introduce
\bea
 \hat{K}_\ka (\al)\bu \defe -i[H_S(\al_S)+\ka H_\RM{L}(\al),\bu]+\sum_b \Pi_b(\al)\bu ,
\eea
and corresponding instantaneous steady state $\rho_0^{(\ka)}(\al)$:
\bea
 \hat{K}_\ka (\al)\rho_0^{(\ka)}(\al)=0. \la{rho_0^ka}
\eea
Here, $\ka$ is a real parameter  satisfying $-1 \le \ka \le 1$ controlling the Lamb shift Hamiltonian. 
$\dbr{1} \hat{K}_\ka (\al) =0$ holds.
We use the following notations:
\bea
\al_{1,b} \defe \be_b, \ \al_{2,b} \defe \be_b\mu_b,\ \overline{X} \defe X \bv{\al_{i,b}=\overline{\al_{i}}} .
\eea

We expand $\rho_0^{(\ka)}$ and $l_0^\pr$ (the derivative of $n=0$ left eigenavector for $\kappa=+1$)
\bea
\rho_0^{(\ka)}(\al) \aeq \overline{\rho_0^{(\ka)}}+\sum_b\left(\ep_{1,b}\rho_{1,b}^{(\ka)}+\ep_{2,b}\rho_{2,b}^{(\ka)}\right)+\mO(\ep^2),\\
l_0^\pr(\al) \aeq \overline{l_0^\pr(\al)}+\sum_b\left(\ep_{1,b}k_{1,b}+\ep_{2,b}k_{2,b}\right)+\mO(\ep^2), \la{l_ep}
\eea
with 
\bea
\overline{\rho_0^{(\ka)}} = \rho_\RM{gc},\ \ \ 
\overline{l_0^\pr(\al) } =-\overline{\be}H_S+\overline{\be\mu}N_S+\overline{c}^\ast1 = \ln \rho_\RM{gc}+\overline{c^\pr}^\ast 1.
\eea
Here, $\rho_\RM{gc}\defe \rho_\RM{gc}(\al_S;\overline{\be},\overline{\be \mu})$, $\overline{c}$ and $\overline{c^\pr}$ are the same with $c(\al)$ and $c^\pr(\al)$ in \res{eq}.
After some calculations, we obtain following relation ($i=1,2$):
\bea
k_{i,b} = \rho_{i,b}^{(-1)}\rho_\RM{gc}^{-1}+\overline{c_{i,b}}1,
\eea
where $\overline{c_{i,b}}$ is an arbitrary complex number.
The details of the derivation are explained in the Appendix \ref{Appendix B}.
Using this relation, \re{l_ep} becomes
\bea
l_0^\pr(\al) 
\aeq \ln \rho_\RM{gc}(\al_S;\overline{\be},\overline{\be \mu})+C(\al)1+\sum_b \sum_{i=1}^2\ep_{i,b}\rho_{i,b}^{(-1)}\rho_\RM{gc}^{-1}+\mO(\ep^2) \no\\
\aeq \ln \rho_0^{(-1)}(\al)+C(\al)1+\mO(\ep^2), \la{Goal-} 
\eea
where $C(\al)\defe\overline{c^\pr}^\ast+\sum_{b,i}\overline{c_{i,b}}\ep_{i,b}$.
Substituting this equation into \re{A_n^sig_F}, we obtain 
\bea
A_n^\sig(\al) = -\tr_S\Big[\ln \rho_0^{(-1)}(\al)\f{\partial \rho_0^{(1)}(\al)}{\partial \al^n}\Big]+\mO(\ep^2), \la{goal}
\eea
where the notation $\rho_0(\al)=\rho_0^{(1)}(\al)$ and $\breve{\rho}_0(\al)=\rho_0^{(-1)}(\al)$ is used in \res{Introduction} for clarity.
We supposed $[\rho_\RM{gc},\rho_{i,b}^{(-1)}] =0$, which leads 
\bea
\ln \rho_0^{(-1)}(\al) =\ln\rho_\RM{gc}+\sum_{i,b}\ep_{i,b}\rho_{i,b}^{(-1)}\rho_\RM{gc}^{-1}+\mO(\ep^2).
\eea
This supposition is satisfied if $[N_S,\rho_0^{(-1)}(\al)]=\mO(\ep^2)$ (which leads $[N_S,\rho_{i,b}^{(-1)}] =0$) or $\overline{\be\mu}=0$ holds. 
If $H_S$ is non-degenerate, $[N_S,\rho_0^{(-1)}(\al)]=0$ holds, then $[N_S,\rho_{i,b}^{(-1)}] =0$, $[\rho_\RM{gc},\rho_{i,b}^{(-1)}] =0$ and \re{Goal-} hold.
If the states of the baths are in the canonical distributions $(\mu_b \to 0)$, $\rho_\RM{gc}$ is replaced by the canonical distribution and 
\re{Goal-} holds without any assumption. 

If 
\bea
[H_\RM{L}(\al), \rho_0^{(\ka)}(\al)]=0, \la{Key2a}
\eea  
holds, $\rho_0^{(\ka)}(\al)$ is independent of $\ka$ ($\rho_0^{(\ka)}(\al)=\rho_0(\al)$), 
then \re{goal} becomes
\bea
A_n^\sig(\al) = \f{\partial }{\partial \al^n} S_\RM{vN}(\rho_0(\al))+\mO(\ep^2), \la{A_S}
\eea
using \re{dS}. 
\re{Key2a} holds if $H_S$ is non-degenerate.
\re{A_S} can be shown from $[\overline{H_\RM{L}},\rho_{i,b}^{(1)}]=0$, which is weaker assumption than \re{Key2a} and is derived from \re{Key2a} for $\ka=1$.
If we neglect the Lamb shift Hamiltonian, namely we consider the QME for $\hat{K}_0(\al)$, \re{A_S} holds (with a replacement $\rho_0 \to \rho_0^{(0)}$).
From \re{A_S}, we obtain 
\bea
\sig_\RM{ex}= S_\RM{vN}(\rho_0(\al_\tau))-S_\RM{vN}(\rho_0(\al_0))+\mO(\ep^2\dl), \la{ex=}
\eea
with 
\bea 
\dl=\max_{n,\al \in C} \f{\abs{\al^n-\al^n_0}}{\abs{\bar{\al}^n}},
\eea
where $\bar{\al}^n$ is typical value of the $n$-th control parameter.

Yuge {\it et al.} \cite{Yuge13} applied the FCS-QME approach to the excess entropy production of the quantum system.
They introduced a {\it time-dependent observable} $A(t)=-\sum_b \be_b(t)[H_b-\mu_b(t)N_b]$ and considered 
the outputs at $t=0$ and $t=\tau$ as $a(0)$ and $a(\tau)$.
Then, they identified the average $\sig^\pr \defe \bra a(\tau)-a(0)\ket$ as the average entropy production. 
However, $\sig^\pr$ seems not the average entropy production $\sig$. 
The average $\sig^\pr $ can be rewritten as 
\bea
\sig^\pr &\approx& \tr_\tot[A(\tau)\rho_\tot(\tau)]-\tr_\tot[A(0)\rho_\tot(0)]
=\int_0^\tau\ dt \left\{\frac{d}{dt}\tr_\tot[A(t)\rho_\tot(t)]\right\}\no\\
\aeqap -\int_0^\tau dt \ \sum_b \Big[\f{d\be_b(t)}{dt}\bra H_b \ket_t-\f{d[\be_b(t)\mu_b(t)]}{dt}\bra N_b \ket_t  \Big] \no\\
&&+\int_0^\tau dt \ \sum_b \Big[\be_b(t)\{ - \f{d}{dt} \bra H_b \ket_t\}-\be_b(t)\mu_b(t)\{ - \f{d}{dt} \bra N_b \ket_t\} \Big]. \la{sig^pr}
\eea
Here, $\bra \bu \ket_t \defe \tr_\tot[\bu \rho_\tot(t)]$, $\rho_\tot(t)$ is the total system state and $\tr_\tot$ denotes the trace of the total system. 
The integrand of the second term of the last expression of \re{sig^pr} roughly equals to $\dot{\sig}$ \footnote{
Here, we supposed $\f{d}{dt}\bra O \ket_t \approx i^O(t)$ for $O=H_b,N_b$. 
However, because  the thermodynamic parameters $\be_b$ and $\mu_b$ are modulated, $\f{d}{dt}\bra H_b \ket_t$ and $\f{d}{dt}\bra N_b \ket_t$ also include 
the currents from the outside of the total system to the bath $b$.}
The first term, while its physical meaning is not clear, is nonzero in general.
Moreover, it should be noted that the FCS-QME is applicable only for a {\it time-independent} observable although $A(t)$ is time-dependent.
These two issues are the problems of Ref. \cite{Yuge13}.
Nevertheless, the obtained Liouvillian (of which the Lamb shift Hamiltonian is neglected) incidentally satisfies \re{G_con}. 
Using that Liouvillian, for the system with time-reversal symmetry,
Yuge {\it et al.} studied the relation between $A_n^\sig(\al)$ and the symmetrized von Neumann entropy.
In contrast, we do not suppose the time-reversal symmetry to derive (\ref{goal}).
In \res{Discussion}, we consider the time-reversal symmetric system.

\subsection{Born-Markov approximation} \la{Born_Markov}

We denote the BSN vector for the entropy production and instantaneous steady state of the Born-Markov approximation by 
$A_n^{\sig,\rm{BM}}(\al)$ and $\rho_0^\RM{BM}(\al)$. 
Then, 
\bea
A_n^{\sig,\rm{BM}}(\al) \aeq A_n^{\sig}(\al)+\mO(v^2),\\
S_\RM{vN}(\rho_0^\RM{BM}(\al)) \aeq S_\RM{vN}(\rho_0(\al)) +\mO(v^2),
\eea
hold \cite{Yuge13}. 
Here, $v=u^2$ and $u(\ll 1)$ describes the order of $H_{Sb}$.
The above two equations and \re{A_S} lead
\bea
A_n^{\sig,\rm{BM}}(\al) = \f{\p }{\p \al^n}S_\RM{vN}(\rho_0^\RM{BM}(\al))+\mO(\ep^2)+\mO(v^2).
\eea

\section{Comparison of two definitions of entropy production} \la{Other_def}

In this section, we compare the preceding study on the entropy production 
in the classical Markov jump process \cite{Komatsu15,Jarzynski} with ours.
We consider the Markov jump process among the states $n=1,2,\cdots,\mN$, where the definitions are explained in Appendix \ref{Appendix C}.
The probability to find the system in a state $n$ is $p_n(t)$ and it obeys the master equation:
\bea
\f{dp_n(t)}{dt} = \sum_{m=1}^\mN K_{nm}(\al_t)p_m(t) .\la{M}
\eea
The Liouvillian is given by 
\bea
K_{nm}(\al)=\sum_b K_{nm}^{(b)}(\al)
\eea
 where $K_{nm}^{(b)}$ originates the couping between the system and the bath $b$.
$ \sum_n K_{nm}^{(b)}(\al)=0$ holds. 
We suppose that $K_{mn}^{(b)}(\al)\ne 0(=0)$ holds if $K_{nm}^{(b)}(\al)\ne0(=0)$ for all $n \ne m$.
The definition of the entropy production for each Markov jump process \re{MJ} is \re{def_EP}. 
The average entropy production $\sig^\RM{C}$ is given by (see \re{sig^C0})
\bea
\sig^\RM{C}  = \int_0^\tau dt \ \sum_{n,m} \sig_{nm}^\RM{C}(\al_t)p_m(t), \la{sig^C}
\eea
where
\bea
\sig_{nm}^\RM{C}(\al)  =  -K_{nm}(\al) \ln \f{K_{nm}(\al)}{K_{mn}(\al)}.
\eea

We denote the solution of the QME with RWA by $\rho(t)$.
We suppose $p_n(t)\defe \br{n}\rho(t)\ke{n}$ is governed by \re{M} with $K_{nm}^{(b)}(\al)=(\Pi_b(\al))_{nn,mm}$. 
Here, $\ke{n}$ is the energy eigenstate of $H_S(\al_S)$, 
\bea
(\Pi_b \bu)_{nm}=\sum_{k,l}(\Pi_b(\al))_{nm,kl} (\bu)_{kl}
\eea
 and $(\bu)_{kl}\defe \br{k}\bu \ke{n}$. 
This supposition implies \re{Key2a}. 
A sufficient condition by which $p_n(t)$ obeys \re{M} is below: 
(1) $H_S(\al_S)$ is non-degenerate and (2) $ \{\al^n \in \al_S \vert \ \f{\p }{\p \al^n}\ke{n} \ne 0 \}$ are fixed. 
The eigenenergy can depend on $\{\al^n \in \al_S \vert \ \f{\p }{\p \al^n}\ke{n} =0 \}$.
We show that our average entropy production \re{def_sig} is given by a similar expression of \re{sig^C}:
\bea
\sig = \int_0^\tau dt \ \sum_{n,m} \sig_{nm}(\al_t)p_m(t). \la{sig^Q}
\eea
Here, 
\bea
 \sig_{nm}(\al) \defe \sum_b K_{nm}^{(b)}(\al)\theta_{nm}^{(b)}(\al) = -\sum_b K_{nm}^{(b)}(\al)\ln \f{K_{nm}^{(b)}(\al)}{K_{mn}^{(b)}(\al)},
\eea
with 
\bea
\theta_{nm}^{(b)}(\al)\defe  \left \{ \begin{array}{ll}
- \ln\f{K_{nm}^{(b)}(\al)}{K_{mn}^{(b)}(\al)} \hs{3mm} K_{nm}^{(b)}(\al) \ne 0 \\
0 \hs{17mm} K_{nm}^{(b)}(\al) = 0
\end{array} \right. .
\eea
Because of \re{def_w}, \re{w=W} and \re{w=W2}, the particle and energy currents are given by
 $i^{X_b}= \tr_S[W^{X_b}\rho(t)]$ with $W^{X_b} = -(\Pi_b \dg X_S)\dg \ (X=H,N) $. 
\re{W_A_B} leads 
\bea
(W^{X_b})_{nm} = -\sum_{k,l} (\Pi_b)_{lk,mn} (X_S)_{kl}.
\eea
We suppose $(X_S)_{nm}=(X_S)_{nn}\dl_{nm}$ for $X=N,H$. 
Since $(X_S)_{kl}$ is a diagonal matrix, $(W^{X_b})_{nm}$ is also a diagonal matrix. 
Then, 
\bea
i^{X_b} = \sum_m (W^{X_b})_{mm}p_m(t), \la{i^X_b3}
\eea
holds. 
Substituting $(W^{X_b})_{mm} =-\sum_n K^{(b)}_{nm} (X_S)_{nn}$ into \re{i^X_b3}, we obtain
\bea
i^{X_b}\aeq -\sum_{n,m}  K^{(b)}_{nm} (X_S)_{nn}p_m(t) \no\\
\aeq \sum_{n,m}  K^{(b)}_{nm} \left[(X_S)_{mm}-(X_S)_{nn}\right]p_m(t).
\eea
This equation leads
\bea
\dot{\sig}(t) = -\sum_{n,m}\sum_b K^{(b)}_{nm} \be_b(t) \{[(H_S)_{mm}-(H_S)_{nn}]-\mu_b(t)[(N_S)_{mm}-(N_S)_{nn}] \}p_m(t).
\eea
Using the local detailed balance condition
\bea
\ln \f{K_{nm}^{(b)}(\al)}{K_{mn}^{(b)}(\al)} = \be_b \{[(H_S)_{mm}-(H_S)_{nn}]-\mu_b[(N_S)_{mm}-(N_S)_{nn}]\},
\eea
we obtain \re{sig^Q}.

Now we introduce a matrix $\mK^\lm(\al)$ by
\bea
[\mK^\lm(\al)]_{nm} \defe \sum_b K_{nm}^{(b)}(\al)e^{i\lm \theta_{nm}^{(b)}(\al)}.
\eea
Then, we obtain
\bea
\f{\partial}{\partial (i\lm)}\Big \vert_{\lm=0} \sum_{n,m}  \Big[ {\rm{T}} \exp \big[ \int_0^\tau dt \ \mK^\lm(\al_t) \big] \Big]_{nm} p_{m}(0) 
= \int_0^\tau dt \ \sum_{n,m}\sig_{nm}(\al_t)p_m(t)=\sig .
\eea
$\mK^\lm$ was originally introduced by Sagawa and Hayakawa \cite{Sagawa}.
About averages, our entropy production is the same with Sagawa and Hayakawa. 

We show that the difference between $\sig^\RM{C}_{nm}(\al)$ and $\sig_{nm}(\al)$ is $\mO(\ep^2)$:
\bea
\sig^\RM{C}_{nm}(\al) = \sig_{nm}(\al)+\mO(\ep^2). \la{sig_sig^C}
\eea
In fact, $K^{(b)}_{nm}$ can be expanded as 
\bea
K_{nm}^{(b)}=\ga_b\bar{K}_{nm}+\sum_{i=1,2} \ep_{i,b}K^{i,b}_{nm}+\mO(\ep^2) ,\ \ \ \sum_b \ga_b = 1,
\eea
then we obtain
\bea
\sig^\RM{C}_{nm}(\al) \aeq \sig_{nm}^{(0,1)}+\sig^{\rm{C}(2)}_{nm}(\al)+\mO(\ep^3) ,\\
 \sig_{nm}(\al) \aeq \sig_{nm}^{(0,1)}+\sig^{(2)}_{nm}(\al)+\mO(\ep^3) ,
\eea
with
\bea
\sig_{nm}^{(0,1)} \defe -\bar{K}_{nm} \ln \f{\bar{K}_{nm}}{\bar{K}_{mn}}+\sum_{i,b}\ep_{i,b} \Big[K^{i,b}_{nm} \ln \f{\bar{K}_{nm}}{\bar{K}_{mn}}+
 K^{i,b}_{nm}-K^{i,b}_{mn}\f{\bar{K}_{nm}}{\bar{K}_{mn}}  \Big].
\eea
$\sig^{\rm{C}(2)}_{nm}(\al)$ and $\sig^{(2)}_{nm}(\al)$ are quadratic orders of $\ep_{i,b}$. 
While the former includes $\ep_{i,b}\ep_{i^\pr,b^\pr}$ ($b \ne b^\pr$) terms, the latter dose not. 
\re{sig_sig^C} leads 
\bea
\sig_\RM{ex}^\RM{C} \aeq \sig_\RM{ex}+\mO(\ep^2\dl).
\eea
Here, $\sig_\RM{ex}^\RM{C}$ is given by \re{sig_ex^C}. 
Then, \re{K15}, the result of Ref. \cite{Komatsu15}, coincides with \re{ex=} when $p_n(t)= \br{n}\rho(t)\ke{n}$ is governed by the master equation \re{M}.

\section{Time-reversal operations} \la{Discussion}

In this section, we define the time-reversal operation and examine the dependence of the excess entropy production on the time-reversal symmetry.
We denote the time-reversal operator of the system by $\theta$. 
We then define 
\bea
\tl Y \defe \theta Y \theta^{-1},
\eea
for all $Y \in {\rm\bm{B}}$ and 
\bea
\tl \mJ \tl Y \defe \theta ( \mJ Y) \theta^{-1},
\eea
for a superoperator $\mJ$ of the system. 
The time-reversal of $\hat{K}(\al)\rho_0(\al)=0$ is given by
\bea
i[\tl H_\RM{L}(\al),\tl \rho_0(\al)]+\sum_b \tl \Pi_b(\al) \tl \rho_0(\al)=0,
\eea
using \re{H_S,rho_0}. 
If 
\bea
\tl H_\RM{L}(\al) =H_\RM{L}(\al),\ \ \tl \Pi_b(\al)= \Pi_b(\al), \la{TRS}
\eea
hold, the above equation coincides with the equation of $\rho_0^{(-1)}(\al)$ since $[H_S,\rho_0^{(\ka)}]=0$,
then 
\bea
\tl \rho_0(\al)=\rho_0^{(-1)}(\al)\defe \breve{\rho}_0(\al), \la{tl=-1}
\eea
holds. 
If the total Hamiltonian is time-reversal invariant, \re{TRS} holds \cite{trs}. 
If \re{TRS} holds and we neglect the Lamb shift Hamiltonian, the instantaneous steady state is time-reversal invariant: $\tl \rho_0^{(0)}=\rho_0^{(0)}$.

For time-reversal symmetric system, 
\bea
\f{\p}{\p \al^n}S_\RM{sym}(\rho_0(\al)) =-\tr_S\Big[\ln \tl \rho_0(\al)\f{\partial \rho_0(\al)}{\partial \al^n}\Big]+\mO(\ep^2), \la{giro}
\eea
holds. 
Here,
\bea
S_\RM{sym}(\rho) \defe - \tr_S\big[\rho\half(\ln \rho+\ln \tl \rho)\big],
\eea
is the symmetrized von Neumann entropy. 
Combining \re{goal} with \re{tl=-1}, we obtain 
\bea
A_n^\sig(\al)=\f{\p}{\p \al^n}S_\RM{sym}(\rho_0(\al))+\mO(\ep^2),
\eea
then, the equation \re{ex=} with $S_\RM{vN} \to S_\RM{sym}$ holds.
As analogy, we consider
\bea
S^\pr(\al) \defe - \tr_S\big[\rho_0(\al)\half(\ln \rho_0(\al)+\ln \breve{\rho}_0(\al))\big],
\eea
for generally non-time-reversal symmetric system.
The difference between $\p S^\pr(\al)/\p \al^n$ and the first term of the RHS of \re{goal} is 
\bea  
&&\f{\p S^\pr(\al)}{\p \al^n}-\Big(-\tr_S\Big[\ln \breve{\rho}_0(\al)\f{\partial \rho_0(\al)}{\partial \al^n}\Big]\Big)\no\\
\aeq -\half \tr_S\big[\f{\p \rho_0}{\p \al^n}(\ln \rho_0-\ln \breve{\rho}_0) \big]-\half  \tr_S\big[\rho_0\f{\p }{\p \al^n}\ln \breve{\rho}_0\big]. \la{giron}
\eea
To calculate the RHS of this equation, we use formulas
\bea
\ln(A+\eta B) \aeq \ln A+\int_0^\infty ds \ \Big(\eta \f{1}{A+s}B\f{1}{A+s}- \eta^2 \f{1}{A+s}B\f{1}{A+s}B\f{1}{A+s}+\mO(\eta^3) \Big) ,\\
\f{\p}{\p \al^n}\ln A(\al) \aeq \int_0^\infty ds \  \f{1}{A(\al)+s}\f{\p A(\al)}{\p \al^n}\f{1}{A(\al)+s} \la{d ln A},
\eea
where $A,B,A(\al)\in {\rm\bm{B}}$ and $\eta$ is small real number.
$\rho_0-\breve{\rho}_0=\ep \psi+\mO(\ep^2)$ holds because $\overline{\rho_0^{(\ka)}} = \rho_\RM{gc}(\al_S;\overline{\be},\overline{\be\nu})$. 
Then, the first term of the RHS of \re{giron} is given by
\bea
-\half \tr_S\big[\f{\p \rho_0}{\p \al^n}(\ln \rho_0-\ln \breve{\rho}_0) \big] = 
-\f{\ep}{2}\int_0^\infty ds \  \tr_S\Big[\f{\p \rho_0}{\p \al^n}\f{1}{\breve{\rho}_0+s}\psi\f{1}{\breve{\rho}_0+s} \Big]+\mO(\ep^2) .
\eea
The second term of the RHS of \re{giron} is given by
\bea
-\half  \tr_S\big[\rho_0\f{\p }{\p \al^n}\ln \breve{\rho}_0 \big] \aeq -\half \int_0^\infty ds \  \tr_S\Big[\f{\p \breve{\rho}_0}{\p \al^n}
\f{1}{\breve{\rho}_0+s}(\breve{\rho}_0+\ep \psi)\f{1}{\breve{\rho}_0+s}  \Big] +\mO(\ep^2)\no\\
\aeq -\half  \tr_S\big[\f{\p \breve{\rho}_0}{\p \al^n}\big]\no\\
&&-\f{\ep}{2} \int_0^\infty ds \  \tr_S\Big[\f{\p \breve{\rho}_0}{\p \al^n}
\f{1}{\breve{\rho}_0+s} \psi\f{1}{\breve{\rho}_0+s}  \Big]+\mO(\ep^2) \no\\
\aeq -\f{\ep}{2} \int_0^\infty ds \  \tr_S\Big[\f{\p \breve{\rho}_0}{\p \al^n}
\f{1}{\rho_0+s} \psi\f{1}{\rho_0+s}  \Big]+\mO(\ep^2) \no\\
\aeq  -\f{\ep}{2} \int_0^\infty ds \  \tr_S\Big[\f{\p (\theta \breve{\rho}_0\theta^{-1})}{\p \al^n}
\f{1}{\tl \rho_0+s} \tl \psi\f{1}{\tl \rho_0+s}  \Big]+\mO(\ep^2).
\eea
Here, we used $\ep(\breve{\rho}_0+s)^{-1}=\ep(\rho_0+s)^{-1}+\mO(\ep^2)$ and $\tr_S\bu=\tr_S \tl \bu$ if $\tr_S \bu$ is real.
In general, the RHS of \re{giron} is not $\mO(\ep^2)$. 
However, if $\tl \rho_0=\breve{\rho}_0$ holds, the RHS of \re{giron} becomes $\mO(\ep^2)$ since $\tl \psi=-\psi$, then \re{giro} holds.
In the proof of \re{giro}, Yuge {\it et al.} \cite{Yuge13} used incorrect equations
$\f{\p}{\p \al^n}\ln \tl \rho_0=\tl \rho_0^{-1}\f{\p\tl \rho_0}{\p \al^n}$ and $\ln \rho_0-\ln \tl \rho_0=\ep \psi \tl \rho_0^{-1}+\mO(\ep^2)$. 

\section{Summary} \la{Summary}

In this paper, for open systems described by the quantum master equation (QME), we investigated the excess entropy production 
under quasistatic operations between nonequilibrium steady states (NESSs). 
We propose a new definition of the average entropy production rate $\dot{\sig}(t)$ using the average energy and particle currents,
which are calculated by using the full counting statistics (FCS) with QME (FCS-QME). 
Then, we introduced the generalized QMEs (GQMEs) providing $\dot{\sig}(t)$. 
The GQMEs do not relate the higher moments (thus and the FCS) of the entropy production, but  
we can calculate only the average of the entropy production. 
Using the GQME, in weakly nonequilibrium regime, we analyzed the Berry-Sinitsyn-Nemenman (BSN) vector for the entropy production, $A_n^\sig(\al)$, 
which provides the excess entropy production $\sig_\RM{ex}$ under quasistatic operations between NESSs as the line integral of
$A_n^\sig(\al)$ in the parameter space.
We have shown that the BSN vector $A_n^\sig(\al)$ for the entropy production is given by
$\rho_0(\al)$, the instantaneous steady state of the QME and $\breve{\rho}_0(\al)$, that of the QME which is given by reversing the sign of the Lamb shift term. 
If the system Hamiltonian is non-degenerate or the Lamb shift term is negligible, we obtain 
that the excess entropy production is given by the difference of the von Neumann entropies at the initial and final times of the operation.
In general, the potential $\mS(\al)$ such that $A_n^\sig(\al)=\f{\p \mS(\al)}{\p \al^n} +\mO(\ep^2)$ dose not exist,
but for time-reversal symmetric system, we showed that $\mS(\al)$ is the symmetrized von Neumann entropy.  
Additionally, we pointed out that preceding expression of the entropy production in the classical Markov jump process \cite{Komatsu15,Jarzynski} is different from ours 
and showed that these approximately equivalent in the weakly nonequilibrium regime. 
We also checked that the definition of the average entropy production in the classical Markov jump process by Ref. \cite{Sagawa} is equivalent to ours.

\renewcommand{\abstractname}{Acknowledgements}
\begin{abstract}
We acknowledge helpful discussions with S. Okada. Part of this work is supported by JSPS KAKENHI (26247051).
\end{abstract}

\appendix

\section{Derivative of the von Neumann entropy} \la{Appendix A}

We show that 
\bea
\f{\p S_\RM{vN}(\rho_0(\al))}{\p \al^n}=-\tr_S \Big[\ln \rho_0(\al) \f{\p \rho_0(\al)}{\p \al^n} \Big] . \la{dS}
\eea
From the definition of the von Neumann entropy, the LHS of the above equation is given by
\bea
\f{\p S_\RM{vN}(\rho_0(\al))}{\p \al^n}=-\tr_S \Big[\ln \rho_0(\al) \f{\p \rho_0(\al)}{\p \al^n} \Big]-\tr_S \Big[ \f{\p \ln \rho_0(\al)}{\p \al^n} \rho_0(\al) \Big].
\eea
Using \re{d ln A}, the second term of the RHS of the above equation becomes
\bea
-\tr_S \Big[ \f{\p \ln \rho_0(\al)}{\p \al^n} \rho_0(\al) \Big] \aeq 
-   \tr_S \Big[ \int_0^\infty ds \ \f{1}{\rho_0(\al)+s}\f{\p \rho_0(\al)}{\p \al^n}\f{1}{\rho_0(\al)+s} \rho_0(\al) \Big] \no\\
\aeq -  \tr_S \Big[ \int_0^\infty ds \ \f{\rho_0(\al)}{(\rho_0(\al)+s)^2}\f{\p \rho_0(\al)}{\p \al^n}  \Big] \no\\
\aeq - \tr_S \Big[ \f{\p \rho_0(\al)}{\p \al^n}  \Big]=0.
\eea
Then, we obtain \re{dS}.

\section{Derivation of the relation between $k_{i,b}$ and $\rho_{i,b}^{\kappa}$} \la{Appendix B}

In this section, we examin the relation of the coefficients of the expansion of $\rho_0^{\kappa}(\al)$ 
and $l_0^\pr(\al)$ in (\ref{l_ep}) of \res{noneq}.

First, we investigate $k_{i,b}$ in \re{l_ep}.
\re{ee} can be rewritten as
\bea
\hat{K}\dg(\al)l_0^\pr(\al)+[\mK^\pr(\al)]\dg1 = J^\sig_\st(\al). \la{eer}
\eea
Here, $J^\sig_\st(\al) = \mO(\ep^2)$ holds
because $i_\st^{H_b}(\al),i_\st^{N_b}(\al)=\mO(\ep)$ and
\bea
J_\st^\sig(\al)=\sum_b \left(-i_\st^{H_b}(\al)\ep_{1,b}+i_\st^{N_b}(\al)\ep_{2,b}\right)
\eea
since
\bea
\sum_b i_\st^{X_b}(\al)= -\tr_S[X_S \sum_b \mL_b(\al) \rho_0(\al)]=0,\ \ \  (X=N,H).
\eea
Then we obtain 
\bea
\overline{\partial_{i,b}\mK^\pr} \dg 1+\overline{K}\dg k_{i,b}+\overline{\partial_{i,b}\mL_{b}} \dg \overline{l_0^\pr} \aeq 0, \la{ee,1}
\eea
in $\mO(\ep_{i,b})$. 
Here, $\partial_{i,b}X \defe \p X/\p \al_{i,b}$ and $\overline{K}\defe \overline{\hat{K}}$.
The first term of the LHS is 
\bea
\overline{\partial_{i,b}\mK^\pr} \dg 1 \aeq  \f{\partial [\mK^\pr]\dg 1}{\partial \al_{i,b}}\Bv{\al_{i,b}=\overline{\al_i}} \no\\
\aeq \f{\partial \mL_b\dg [\al_{1,b}H_S-\al_{2,b}N_S]}{\partial \al_{i,b}}\Bv{\al_{i,b}=\overline{\al_i}} \no\\
\aeq \overline{\partial_{i,b} \mL_b} \dg[\overline{\be}H_S-\overline{\be \mu}N_S]+\overline{\Pi_b}\dg \f{\partial[\al_{1,b}H_S-\al_{2,b}N_S]}{\partial \al_{i,b}}.
\eea
The third term of the LHS becomes
\bea
\overline{\partial_{i,b}\mL_{b}} \dg \overline{l_0^\pr} \aeq \overline{\partial_{i,b}\mL_{b}} \dg ( -\overline{\be}H_S+\overline{\be\mu}N_S+\overline{c^\pr}^\ast 1) \no\\
\aeq -\overline{\partial_{i,b}\mL_{b}} \dg(\overline{\be}H_S-\overline{\be\mu}N_S).
\eea
Here, we used $\overline{\partial_{i,b}\mL_{b}}\dg1=0$ derived from $\hat{K}\dg 1=0$. 
Then, \re{ee,1} becomes
\bea
\overline{K}\dg k_{1,b}+\overline{\Pi_b}\dg H_S \aeq 0,\la{ee,2} \\
\overline{K}\dg k_{2,b}-\overline{\Pi_b}\dg N_S \aeq 0 \la{ee,3}.
\eea
Next, we show the relation between $k_{i,b}$ and $\rho_{i,b}^{(-1)}$. 
\re{rho_0^ka} leads 
\bea
\overline{K_\ka}\rho_{i,b}^{(\ka)}+\overline{\partial_{i,b}\mL_b}\rho_\RM{gc} = 0 , \la{ss,1}
\eea
in $\mO(\ep_{i,b})$. 
Here, $\overline{K_\ka}\defe \overline{\hat{K}}_\ka$.
By the way,
\bea
\mL_b \rho_\RM{gc}(\al_S;\be_b,\be_b \mu_b)=0,
\eea
holds. 
Differentiating this equation by $\al_{i,b}$, we obtain
\bea
\overline{\partial_{i,b}\mL_b}\rho_\RM{gc} \aeq -\overline{\mL_b}\overline{\f{\rho_\RM{gc}(\al_S;\be_b,\be_b \mu_b)}{\partial \al_{i,b}}} =
 \overline{\mL_b}\f{\partial [\al_{1,b} H_S-\al_{2,b}N_S]}{\partial \al_{i,b}}\rho_\RM{gc}(\al_S;\overline{\be},\overline{\be \mu}).
\eea
Substituting these equations into \re{ss,1}, we obtain
\bea
\overline{K}_\ka \rho_{1,b}^{(\ka)}+\overline{\Pi_b}(H_S \rho_\RM{gc}) \aeq 0, \la{ss,1c}\\
\overline{K}_\ka \rho_{2,b}^{(\ka)}-\overline{\Pi_b}(N_S \rho_\RM{gc}) \aeq 0.\la{ss,1d}
\eea
Now, we use 
\bea
\overline{\Pi_b}(\bu\rho_\RM{gc})=(\overline{\Pi_b}\dg \bu)\rho_\RM{gc} \la{DB2m},
\eea
which is derived from KMS condition \re{KMS}. 
Using this relation, we rewire \re{ss,1c} and \re{ss,1d} as
\bea
\overline{K}_\ka\rho_{1,b}^{(\ka)}+(\overline{\Pi_b}\dg H_S) \rho_\RM{gc} \aeq 0, \la{ss,1e}\\
\overline{K}_\ka\rho_{2,b}^{(\ka)}-(\overline{\Pi_b}\dg N_S) \rho_\RM{gc} \aeq 0 \la{ss,1f}.
\eea
Multiplying $\rho_\RM{gc}^{-1}$ from the right, we obtain
\bea
(\overline{K}_\ka\rho_{1,b}^{(\ka)})\rho_\RM{gc}^{-1}+\overline{\Pi_b}\dg H_S\aeq 0, \la{ss,1g}\\
(\overline{K}_\ka\rho_{2,b}^{(\ka)})\rho_\RM{gc}^{-1}-\overline{\Pi_b}\dg N_S\aeq 0 \la{ss,1h}.
\eea
 \re{DB2m} can be rewritten as
\bea
(\overline{\Pi_b}Y)\rho_\RM{gc}^{-1}=\overline{\Pi_b}\dg(Y\rho_\RM{gc}^{-1}), \la{DB2r}
\eea
for any $Y=\bu\rho_\RM{gc} \in {\rm\bm{B}}$ by multiplying $\rho_\RM{gc}^{-1}$ from the right.
\re{DB2r} leads
\bea
(\overline{\Pi}\rho_{i,b}^{(\ka)})\rho_\RM{gc}^{-1} =\overline{\Pi}\dg(\rho_{i,b}^{(\ka)}\rho_\RM{gc}^{-1}) ,\la{K1}
\eea
where $\overline{\Pi}\defe \sum_b \overline{\Pi_b}$.
By the way, $[H_S(\al_S),\rho_0^{(\ka)}(\al)]= 0$ holds similarly to \re{H_S,rho_0}. 
Differentiating this equation by $\al_{i,b}$, we obtain
\bea
[H_S(\al_S),\rho_{i,b}^{(\ka)}] = 0. \la{Key}
\eea
This relation leads 
\bea
(\overline{H_\ka^{\times}}\rho_{i,b}^{(\ka)})\rho_\RM{gc}^{-1}=\overline{H_\ka^{\times}}(\rho_{i,b}^{(\ka)}\rho_\RM{gc}^{-1})
= \overline{H_{-\ka}^{\times}}\dg(\rho_{i,b}^{(\ka)}\rho_\RM{gc}^{-1}), \la{K2}
\eea
where $H_\ka^{\times}\bu \defe -i[H_S(\al_S)+\ka H_\RM{L}(\al),\bu]$. 
We used $(H_\ka^{\times})\dg=-H_\ka^{\times}$. 
In the first equality, we used that $\rho_\RM{gc}$ commutes with $H_S$ and $H_\RM{L}$.
\re{K1} and \re{K2} lead
\bea
(\overline{K}_\ka\rho_{i,b}^{(\ka)})\rho_\RM{gc}^{-1} = \overline{K}_{-\ka}\dg(\rho_{i,b}^{(\ka)}\rho_\RM{gc}^{-1}). \la{imp}
\eea
Substituting this into \re{ss,1g} and \re{ss,1h}, we obtain
\bea
 \overline{K}_{-\ka}\dg (\rho_{1,b}^{(\ka)}\rho_\RM{gc}^{-1})+\overline{\Pi_b}\dg H_S\aeq 0 ,\la{ss,1i}\\
  \overline{K}_{-\ka}\dg (\rho_{2,b}^{(\ka)}\rho_\RM{gc}^{-1})-\overline{\Pi_b}\dg N_S\aeq 0 \la{ss,1j}.
\eea
Subtracting \re{ss,1i} (\re{ss,1j}) for $\ka=-1$ from \re{ee,2} (\re{ee,3}), we obtain
\bea
\overline{K}\dg(k_{i,b}-\rho_{i,b}^{(-1)}\rho_\RM{gc}^{-1}) = 0.
\eea
This means
\bea
k_{i,b} = \rho_{i,b}^{(-1)}\rho_\RM{gc}^{-1}+\overline{c_{i,b}}1,
\eea
where $\overline{c_{i,b}}$ is an arbitrary complex number.

\section{Definition of entropy production of the Markov jump process} \la{Appendix C}

Except \re{GMEn}, this section is based on Ref. \cite{Komatsu15}. 
We consider the Markov jump process on the states $n=1,2,\cdots,\mN$:
\bea
n(t)= n_k \hs{3mm} (t_k \le t<t_{k+1}), \ t_0= 0<t_1<t_2 \cdots <t_n<t_{N+1}=\tau . \la{MJ}
\eea
where $N=0,1,2,\cdots$ is the total number of jumps.
We denote the above path by 
\bea
\hat{n}= (N,(n_0,n_1,\cdots,n_N),(t_1,t_2,\cdots,t_N)) . \la{Hat_n}
\eea
The probability to find the system in a state $n$ is $p_n(t)$ and it obeys the master equation \re{M}. 
We suppose the trajectory of the control $\hat{\al}=\big( \al(t) \big)_{t=0}^{\tau}$ is smooth. 
Now we introduce
\bea
\theta_{nm}(\al) \defe  \left \{ \begin{array}{ll}
- \ln\f{K_{nm}(\al)}{K_{mn}(\al)} \hs{8mm} K_{nm}(\al) \ne 0 \\
0 \hs{20mm} K_{nm}(\al) = 0
\end{array} \right. .
\eea
If $n\ne m$, this is entropy production of process $m \to n$. 
The entropy production of process \re{Hat_n} is defined by 
\bea
\Theta^{\hat{\al}}[\hat{n}] = \sum_{k=1}^N \theta_{n_kn_{k-1}}(\al_{t_k}). \la{def_EP}
\eea
Then the weight (the transition probability density) associated with a path $\hat{n}$ is 
\bea
\mT^{\hat{\al}}[\hat{n}] = \prod_{k=1}^N K_{n_kn_{k-1}}(\al_{t_k}) \exp \Big[\sum_{k=0}^N \int_{t_k}^{t_{k+1}}dt \ K_{n_kn_k}(\al_t) \Big] .
\eea
The integral over all the paths is defined by
\bea
\int \mD \hn \ Y[\hn] \defe \sum_{N=0}^\infty \sum_{n_0,n_1,\cdots,n_N}^{n_{k-1}\ne n_k}\int_0^\tau dt_1 \int_{t_1}^\tau dt_2 \int_{t_3}^\tau dt_3 \cdots
 \int_{t_{N-1}}^\tau dt_N \ Y[\hn],
\eea
and the expectation value of $X[\hn]$ is defined by
\bea
\bra X \ket^\ha \defe \int \mD \hn \ X[\hn]p_{n_0}^\st(\al_0) \mT^\ha[\hn] .
\eea
Here, $p_{n}^\st(\al)$ is the instantaneous stationary probability distribution characterized by $\sum_m K_{nm}(\al)p_m^\st(\al) = 0$. 
We introduce a matrix $K^\lm(\al)$ by
\bea
[K^\lm(\al)]_{nm} \defe K_{nm}(\al)e^{i\lm \theta_{nm}(\al)} .
\eea
Then, the $k$-th order moment of the entropy production is given by 
\bea
\bra (\Theta^\ha[\hn])^k \ket^\ha
= \f{\partial^k}{\partial (i\lm)^k}\Big \vert_{\lm=0} \sum_{n,m}  \Big[ {\rm{T}} \exp \big[ \int_0^\tau dt \ K^\lm(\al_t) \big] \Big]_{nm}   p_m^\st(\al_0) . \la{GMEn}
\eea
In particular, the average is given by 
\bea
\sig^\RM{C} \defe \bra \Theta^\ha[\hn] \ket^\ha = \int_0^\tau dt \ \sum_{n,m} \sig_{nm}^\RM{C}(\al_t)p_m(t), \la{sig^C0}
\eea
where
\bea
\sig_{nm}^\RM{C}(\al) \defe K_{nm}(\al) \theta_{nm}(\al) =  -K_{nm}(\al) \ln \f{K_{nm}(\al)}{K_{mn}(\al)}.
\eea
According to Ref. \cite{Komatsu15}, for a quasi-static operation, 
\bea
\sig_\RM{ex}^\RM{C}= S_\RM{Sh}[p^\st(\al_\tau)]-S_\RM{Sh}[p^\st(\al_0)]+\mO(\ep^2\dl), \la{K15}
\eea
holds where 
\bea
\sig_\RM{ex}^\RM{C}\defe \sig^\RM{C}-\int_0^\tau dt \ \sum_{n,m} \sig_{nm}^\RM{C}(\al_t)p_m^\st(\al_t) , \la{sig_ex^C}
\eea 
and $ S_\RM{Sh}[p]\defe -\sum_n p_n\ln p_n$.


\end{document}